\documentclass[PAPER, atlasdraft=false, cernpreprint, coverpage=false, USenglish, texmf, orcidlogo]{atlasdoc}
\usepackage{atlaspackage}
\usepackage{atlasbiblatex}

\usepackage{atlasphysics}
\graphicspath{{logos/}{figures/}}

\usepackage{ANA-STDM-2019-20-PAPER-defs}

\AtlasTitle{Search for the exclusive $W$ boson hadronic decays $W^{\pm}\to\pi^{\pm}\gamma$,
$W^{\pm}\to K^{\pm}\gamma$ and $W^{\pm}\to\rho^{\pm}\gamma$ with the ATLAS detector}

\AtlasAbstract{%
A search for the exclusive hadronic decays $W^{\pm}\to \pi^{\pm}\gamma$,
$W^{\pm}\to K^{\pm}\gamma$ and $W^{\pm}\to \rho^{\pm}\gamma$ is performed using
up to $140\, \text{fb}^{-1}$ of proton-proton collisions recorded with the ATLAS
detector at a center-of-mass energy of $\sqrt{s}=13\,\text{TeV}$.
If observed, these rare processes would provide an unique test bench for the quantum chromodynamics
factorization formalism used to calculate cross sections at colliders. Additionally, at future colliders, these decays could offer a new way to measure the
$W$ boson mass through fully reconstructed decay products.
The search results in the most stringent upper limits to date on the branching
fractions $\mathcal{B}(W^{\pm}\to \pi^{\pm}\gamma) < 1.9\times10^{-6}$,
$\mathcal{B}(W^{\pm}\to K^{\pm}\gamma) < 1.7\times10^{-6}$,
$\mathcal{B}(W^{\pm}\to \rho^{\pm}\gamma) < 5.2\times10^{-6}$ at 95\% confidence
level.
}

\AtlasRefCode{STDM-2019-20}

\PreprintIdNumber{CERN-EP-2023-211}

\AtlasJournal{PRL}
\AtlasJournalRef{Phys. Rev. Lett. 133 (2024) 161804}
\AtlasDOI{10.1103/PhysRevLett.133.161804}

\hypersetup{pdftitle={ATLAS document},pdfauthor={The ATLAS Collaboration}}

\begin{document}

\maketitle

%

%
%

%
%
%
The $\Wboson$ boson predominantly decays hadronically into a quark-antiquark pair
that manifests as a pair of jets. In rare cases, the quark pair gives rise to
one or a few hadrons. Examples include decays with a meson $M^{\pm}$ and a photon in the final state, of the form $\Wboson^{\pm}\to M^{\pm}\gamma$~\cite{Grossmann:2015lea,Melia:2016knk}, and fully hadronic decays such as $\Wppp$~\cite{Melia:2016knk}.
These decays offer a unique opportunity to study both the high energy/weakly coupled and low energy/strongly coupled regimes of quantum chromodynamics (QCD) in a single process. In particular, radiative decays are a test bench for the QCD factorization
framework~\cite{Lepage:1979zb,Lepage:1980fj,Efremov:1978rn,Efremov:1979qk}
which allows the calculation of cross sections of processes at hadron
colliders through the separation of perturbative and non-perturbative elements.
The importance of these decays in this context arises from the fact that
higher order corrections constitute small contributions, since they scale with $\Lambda_{\text{QCD}}/m_{\Wboson}$~\cite{Grossmann:2015lea}, where $\Lambda_{\text{QCD}}$ denotes the QCD energy scale.
Furthermore, these exclusive decays could be explored as a new way to measure
the $\Wboson$ boson mass~\cite{Jones:2020bvu, Mangano:2014xta}.
The experimental precision of the $\Wboson$ boson mass measurement is currently inferior to that of
the Standard Model prediction~\cite{ATLAS:2024erm}.
This measurement has been performed in the past exclusively through leptonic
decays of the $\Wboson$ boson ($\Wboson\to\ell\nu$), with often large
systematic uncertainties associated with the incomplete kinematics due to the
presence of the neutrino. At future colliders beyond the HL-LHC~\cite{ZurbanoFernandez:2020cco}, the aforementioned exclusive hadronic decays could
enable a $\Wboson$ boson mass measurement with a fully reconstructed, high
resolution final state.

As a result of their importance, there are multiple theoretical predictions
for the branching fractions of these decays~\cite{Arnellos:1981gy, Manohar:1990hu, Keum:1993eb,
Mangano:2014xta, Grossmann:2015lea}.
These span orders of magnitude and experimental input is required to shed light
on this puzzle. To date, these decays have remained largely unexplored, and no
exclusive hadronic decay of the $\Wboson$ boson has been observed.
The most stringent upper limits at \SI{95}{\%} confidence level (CL) are
$\mathcal{B}(\Wpi) < 7.0\times10^{-6}$ by the CDF Collaboration~\cite{Aaltonen:2011nn},
$\mathcal{B}(\WD) < 6.5\times10^{-4}$ by the LHCb Collaboration~\cite{LHCb:2022kta},
and $\mathcal{B}(\Wppp) < 1.01\times10^{-6}$ by the CMS Collaboration~\cite{CMS-SMP-18-009}.
Upper limits on \Wpi have also been published by the UA2 and CMS
Collaborations~\cite{UA2:1991sgc, CMS-SMP-20-008}.

\begin{figure}[htbp]
\centering
\includegraphics[width=0.6\linewidth]{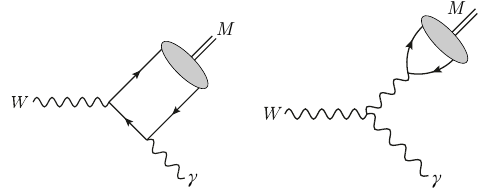}
\caption{Leading-order Feynman diagrams for the radiative decays $\Wboson\to M\gamma$ with
$M = \{\pi, K, \rho\}$. The fermion lines represent quarks, the gray blobs
represent the meson bound state.}
\label{fig:feyn}
\end{figure}

This Letter reports a search for \Wpi, \WK and \Wrho.
The latter two decays have not been previously searched for by other experiments.
The leading-order Feynman diagrams representing the three decay processes are
shown in \cref{fig:feyn}.
The most recent predictions for the branching fractions are
$(4.0 \pm 0.8) \times 10^{-9}$,
$(3.3 \pm 0.7) \times 10^{-10}$, and
$(8.7 \pm 1.9) \times 10^{-9}$, respectively~\cite{Grossmann:2015lea}.
The analysis presented here uses up to \SI{140}{\ifb} of proton-proton ($pp$)
collision data at the center-of-mass energy of $\sqrt{s} = \SI{13}{\TeV}$
collected by the ATLAS experiment between 2015 and 2018.
The analysis is enabled by novel experimental techniques: a dedicated trigger
targeting final states with a single hadron, a non-parametric background
modeling method, and the unconventional use of photon triggers and
$\tau$-lepton reconstruction algorithms to target the $\rho^{\pm}\to\pi^{\pm}\pi^{0}$
decay. In fact, a $\tau$ lepton decays into $\pi^{\pm}\pi^{0}$, via
an intermediate $\rho^{\pm}$, $25.5\%$ of the time~\cite{PDG:2022}.
The limited ATLAS particle identification capabilities for high momentum hadrons
do not allow discrimination between \WK and \Wpi.
The two processes are collectively referred to as $\WpiK$ in the following and
distinguished when necessary.

ATLAS~\cite{PERF-2007-01} is a multipurpose particle detector at the LHC with
cylindrical geometry and a near \(4\pi\) coverage in solid angle.
It consists of an inner tracking detector surrounded by a superconducting
solenoid, electromagnetic and hadronic calorimeters, and a muon spectrometer
with three toroidal superconducting magnets.
The inner tracking detector (ID) covers the pseudorapidity range \(|\eta| < 2.5\).
It consists of silicon pixel, silicon microstrip, and transition radiation
tracking (TRT) detectors. Liquid-argon (LAr) sampling calorimeters provide
electromagnetic energy measurements with high granularity.
A steel and scintillator tile hadron calorimeter covers the central pseudorapidity
range (\(|\eta| < 1.7\)). The endcap and forward regions are instrumented with
LAr calorimeters for both the electromagnetic and hadronic energy measurements up to
\(|\eta| = 4.9\). A two-level trigger system is used to select events.
The first-level trigger, implemented in hardware, uses a subset of the
detector information to accept events at a rate below \SI{100}{\kHz}.
A software-based trigger, part of the ATLAS software suite~\cite{ATL-SOFT-PUB-2021-001},
further reduces the accepted event rate to \SI{1}{\kHz} on average.

The main backgrounds are multijet and mis-reconstructed \Zee events, in which
one electron is mis-reconstructed as a photon and the other electron is
mis-reconstructed as a meson candidate.
Multijet events are modeled with a data-driven method described in
Ref.~\cite{Chisholm:2021pdn} and employed in previous ATLAS
analyses~\cite{HIGG-2016-05, HIGG-2016-13, HIGG-2016-23, HDBS-2018-53}.
Monte Carlo (MC) simulation is used to model the \Zee and signal processes.
Events are generated at next-to-leading order precision in QCD with
\POWHEGBOX{ v1}~\cite{powheg} using the CT10~\cite{CT10} set of
parton distribution functions (PDFs). The parton shower, hadronization and
underlying event are modeled with \PYTHIA{8}~\cite{pythia}
(version 8.243 for \Wpi, 8.244 for \WK and \Wrho, 8.186 for \Zee)
configured according to the \AZNLO tune~\cite{STDM-2012-23} using the CTEQ6L1
PDF set~\cite{ctlq61}.
In \WM events, the $\W$ boson is decayed isotropically and events are reweighted to
match the theoretically predicted angular distribution~\cite{bookpol}.
For the $\W$ boson production cross section the ATLAS measurement of
\SI{185 \pm 6}{nb} is used~\cite{STDM-2015-03}.
The detector response is simulated with a \GEANT-based~\cite{Agostinelli:2002hh}
ATLAS framework~\cite{SOFT-2010-01}.
The effect of additional interactions in the same and neighboring bunch
crossings (pileup) is modeled by overlaying simulated inelastic $pp$ events
generated by \PYTHIA{8} with the A3 tune~\cite{ATL-PHYS-PUB-2016-017} and the
NNPDF2.3lo PDF set~\cite{Ball:2012cx}. MC events are reweighted so that the
distribution of the average number of interactions per bunch crossing matches
the one in the data.
Only events recorded during stable beam conditions, and for which all relevant
components of the detector were operational, are considered~\cite{DAPR-2018-01}.

Two orthogonal event selections are defined, 1) \textit{track-photon}, optimized for
\Wpi and \WK, reconstructing the charged meson as a track, and 2) \textit{tau-photon}, targeting \Wrho. In the \TAPH selection, the meson
candidate is reconstructed as a hadronic $\tau$ lepton ($\tauhad$) taking into
account both the charged and neutral $\rho^{\pm}$ meson decay products.
The \TRPH selection offers supplementary sensitivity to \Wrho events that are
partially reconstructed, because the $\pi^{0}$ is not explicitly identified.
Two sets of triggers are used to record events: a \TRPH trigger for the \TRPH
selection and a di-photon trigger for the \TAPH selection.
The track-photon trigger was derived from $\tau$-lepton
triggers~\cite{ATLAS-CONF-2017-061} and modified to select \Wpi events.
This trigger was activated in 2016 and collected a dataset of $137\,\ifb$.
It requires a photon with transverse momentum $\pt > 25\,\GeV$
($35\,\GeV$ in 20\% of the dataset) and one isolated ID track with
$\pt > 30\,\GeV$ associated with a topological cluster of calorimeter
cells~\cite{PERF-2014-07} with transverse energy $\ET > 25\,\GeV$.
The invariant mass of the track and photon is required to be greater than
$50\,\GeV$ and the ratio between the energy deposition in the calorimeter
matched to the track and the track transverse momentum $\ET/\pt$ is required
to lie within 0.4 and 0.85 to limit the trigger rate for background processes. This trigger has 58\% efficiency for selecting \Wpi events in the phase space of interest.
While the requirement on $\ET/\pt$ is efficient for \WpiK, it significantly reduces the
acceptance for \Wrho events. Consequently, a di-photon trigger~\cite{TRIG-2016-01} that requires
two photons with $\pt > 35\,\GeV$ and $\pt > 25\,\GeV$, respectively, is
employed to recover sensitivity to $\Wrho$ events.
The di-photon trigger sensitivity to \Wrho derives from the production of two
collimated photons in the decay
chain $\rho^{\pm} \to \pi^{\pm}\pi^{0},\ \pi^{0}\to\gamma\gamma$. The efficiency of this trigger for events selected by the \TAPH selection requirements is 43\%.
The $\pt$ requirements of the analysis are mainly driven by the high $\pt$ thresholds of the triggers, which are necessary in order to limit the high background rate from multijet events.

Tracks are reconstructed from hits in the ID, as described in Ref.~\cite{PERF-2015-08}, and are
required to have $\pt > 33\,\GeV$ and to be within the acceptance of the ID. Tracks must also satisfy the
``Tight Primary'' quality criteria detailed in Ref.~\cite{ATL-PHYS-PUB-2015-051}, in order to reject displaced tracks not directly produced in $pp$ collisions and to reduce backgrounds from random combinations of hits. Furthermore, the sum of transverse momenta of tracks within a
cone of $\dR = \sqrt{\Delta\phi^{2} + \Delta\eta^{2}} = 0.2$, excluding the candidate track, is required to be
less than 14\% of its $\pt$, hence imposing low hadronic activity surrounding the selected meson, and suppressing contributions from tracks associated to jets. The $\pt$ and isolation requirements were optimized for a maximum significance of \Wpi over the multijet background for the subset of data with a \Wboson boson invariant mass between \SI{75}{\GeV} and \SI{85}{\GeV}.

Photon candidates are reconstructed from variable-size topological
clusters in the LAr calorimeter~\cite{ATL-PHYS-PUB-2017-022}. Identification is performed using a multivariate discriminant trained on shower shape variables to reject background from hadronic jets~\cite{EGAM-2018-01}. The ``Tight'' criterion described in Ref.~\cite{EGAM-2018-01} is used. Photons with $|\eta|> 2.37$ are discarded, since they are outside of the acceptance of the first layer of the LAr calorimeter, employed in photon-$\pi^{0}$ discrimination. Photon candidates in the transition region between the barrel and endcap are also excluded, since performance is degraded in this region due to a substantial amount of inactive material in front of the active layers of the calorimeter. In order to be considered for the \TRPH (\TAPH) selection, photons are required to have $\pt > 30\,\GeV$ or $\pt > 35 \, (36)\,\GeV$, depending on the trigger being used. Both selections reduce contributions from jets faking photons by requiring that the sum of transverse energies
of calorimeter clusters not associated to the photon candidate but within a cone of $\dR = 0.4$ is less than $2.45\;\GeV + 0.022 \times \pty$ and the sum of transverse momenta of tracks within a cone of $\dR = 0.2$, excluding possible conversion tracks, is less than 5\% of $ \pty$.

The \tauhad reconstruction~\cite{PERF-2014-06, ATLAS-CONF-2017-029} considers
only visible decay products and is seeded by jets built by combining calibrated
calorimeter clusters.
Charged constituents are reconstructed by matching tracks within $\dR < 0.2$ of the \tauhad direction to calorimeter
clusters using a particle-flow approach, while neutral constituents
($\pi^{0}_{\text{cand}}$) are reconstructed from clusters surrounding the charged
candidates. Mis-reconstructed $\pi^{0}_{\text{cand}}$, arising from
$\pi^{\pm}$ cluster remnants or noise, are rejected using a multivariate discriminant.
Five \tauhad classes are defined according to the number of $\pi^{\pm}$ and $\pi^{0}$
and the migration across classes is mitigated by a second multivariate discriminant.
Mis-reconstructed \tauhad are suppressed using an identification algorithm based
on a recurrent neural network~\cite{ATL-PHYS-PUB-2019-033}.
\tauhad objects with exactly one $\pi^{\pm}$ and one $\pi^{0}$ as constituents are used to reconstruct
$\rho^{\pm}\to\pi^{\pm}\pi^{0}$ candidates in the \Wrho decay.
The \tauhad reconstruction algorithm is well suited for the prompt $\rho^{\pm}$
decay as the latter is indistinguishable from one produced in the $\tau$ decay
$\tau^{\pm} \to \rho^{\pm}\nu_{\tau}$ besides a small
displacement of the $\pi$ track due to the decay length of the $\tau$ lepton.
The candidate $\rho^{\pm}$ meson is required to have $\pt > 30\,\GeV$. On average, a $\rho^{\pm} \to \pi^{\pm}\pi^{0}$ decay will be contained in a narrower angular cone than a hadronic jet, and as such the angular distance $\dR_{\tauhad}$ is required to be less than 0.065. Additionally, a requirement of $\log(|d_{0}|/\si{\milli\metre}) < -1.2$, where $d_{0}$ stands for the transverse impact parameter of the track, is used. This ensures that the charged particle originates from the interaction region, rather than from the decay of a tau lepton or another long lived particle. The specific values of the selections were chosen following simultaneous optimization for the maximum significance of \Wrho over the multijet background.

The \Zee background suppression in the \TRPH selection is based on the expected different behavior
between an electron and a charged hadron in terms of the ratio between the
transverse energy deposited in the hadronic and electromagnetic calorimeters
(hadronic leakage), and the amount of transition radiation generated in the TRT.
The latter one is used to construct a likelihood discriminant employed in
electron versus charged-hadron discrimination, which is defined in Ref.~\cite{ATLAS-CONF-2016-024}.
Typically, electrons are expected to generate larger levels of transition radiation and to have very low
hadronic leakage, compared to charged hadrons. In the \TAPH selection, the \Zee background is suppressed by applying requirements on the properties of the
$\tauhad$ object and its constituents. As in the \TRPH case, the amount of radiation produced in the TRT by the $\tauhad$ track is exploited. $\tauhad$ objects with a low value of $\ET^{\tauhad}/\pt^{\pi^{\pm}}$ are excluded, since mis-reconstructed electrons have, on average, lower $\ET^{\tauhad}$ than $\rho^{\pm}$ candidates. A multivariate discriminant trained on \tauhad calorimeter shower shape variables and track properties to discriminate between hadronic tau lepton decays and electrons~\cite{ATL-PHYS-PUB-2015-045} is employed. Differences in the kinematic topology between a single electron mis-reconstructed as a \tauhad and a $\rho^{\pm}$ decay are also exploited. In the case of the electron, the direction of the \tauhad object is defined solely by the electron, and as such the charged track will be closer in $\dR$ to the \tauhad axis. A lower limit is thus imposed on $\dR_{\tauhad}$.

The decay products of the signal processes are generated back-to-back in the laboratory frame. Therefore, in both selections, events with low $\Delta\phi(M, \gamma)$ are excluded. Furthermore, in the \TRPH selection, if both the meson and the photon are reconstructed in the endcap regions, the photon and meson candidates are required to have $\eta(M) \times \eta(\gamma) \geq 0$. A \Wboson boson candidate in the \TRPH selection is formed by the highest \pt photon and meson, while in the \TAPH selection the meson and photon candidate pair with with the largest $\Delta\phi$ is selected. A detailed list of the selection criteria used is provided in the Supplemental Material of this Letter. The \TRPH selection efficiency (including the trigger selection) is 5.0\% for \Wpi, 5.5\% for
\WK and 0.5\% for \Wrho. The higher \WK efficiency compared to \Wpi originates from the differences in nuclear interaction properties for pions and kaons which create small differences for variables used in the trigger selection (mainly
the track $\ET/\pt$) and in the \Zee background suppression. The efficiency of the \TAPH selection for \Wrho is 0.3\%,
half that of the \TRPH signal region (SR) selection but compensated by a higher background rejection.
The contribution of \Wpi and \WK events surviving the \TAPH selection requirements is negligible, with
a number of predicted events $<10^{-8}$ for each process. The contribution of \Wtaunu events is also negligible, suppressed by the
requirements on the photon including its separation from the $\tauhad$ in the
transverse plane.

The background is dominated by the multijet component, which is modeled
using a non-parametric data-driven technique~\cite{Chisholm:2021pdn}. This method models the most important features of the background in a dataset defined by a relaxed version of the event selection, the Generation Region (GR).
In the \TRPH selection, the GR is defined by relaxing the requirements on the $\pt$ and isolation of the track, as well as the isolation requirements on the photon. In the case of the \TAPH selection, the GR is constructed by relaxing the requirements on $\pt(\tauhad)$, $\dR_{\tauhad}$ and $\log(|d_{0}|)$ of the $\tauhad$ track. Data events in the GR are used to construct templates of the variables needed
to describe the kinematics of the decay products and object properties, such as
isolation. Templates use up to three dimensions to capture the most relevant correlations
across variables. Multijet pseudo-events are generated by ancestral sampling of the templates. Details of the sampling sequences employed can be found in the Supplemental Material of this Letter. Following the sampling procedure, compound kinematic variables, such as the $\Wboson$ boson invariant mass, can be calculated for each generated pseudo-event. The process of factorizing the n-dimensional probability density function of the background into the product of lower dimension distributions dilutes the features of resonant contributions, such as the small \Zee background. These contributions are not described by the model and need to be considered separately. Other resonant processes such as $Z\to\tau\tau$ and $W/Z \to qq$ were found to have a negligible impact on the sensitivity. Possible small contributions from non-resonant background processes are absorbed by the background modelling method.
The set of produced multijet pseudo-events is normalized to the number of
observed data events in GR and it is subject to the full set of selection requirements which define the Signal Region (SR). Validation regions (VRs) are defined by applying only one of the SR requirements to events in the GR. Good compatibility is found between the
data and the background prediction in these VRs, verifying the correct modeling
of the most important correlations in the data sample.

The discriminating variable used to quantify the presence of signal is the
candidate $\W$ boson invariant mass.
In the \TRPH selection, the shapes of the $m(\Wpi)$ and $m(\WK)$ distributions
are modeled with the same functional form, a sum of two Voigt functions
multiplied by a sigmoid-like efficiency curve obtained by fitting MC event
distributions in the SR. A single Voigt function is used for $m(\Wrho)$ in the
\TAPH selection as no goodness-of-fit improvement is observed by using two.
The efficiency curve describes the variation of acceptance as a function of the
candidate $\W$ boson invariant mass.
In the \TRPH selection, the $m(\Wrho)$ shape is obtained by smoothing the MC
events with a Gaussian kernel density estimator (KDE).
The resulting $\W$ boson mass resolution is $2.7\%$ for \WpiK; $3.1\%$ for \Wrho
in the \TRPH selection, and $2.9\%$ in the \TAPH selection.
In both selections, the multijet and \Zee predictions are also KDE-smoothed.

The presence of a signal is quantified using a binned maximum likelihood fit.
The mass range between 60 \GeV and 110 \GeV is used for both selections.
The background normalization is determined in the fit.
Systematic uncertainties associated with the shape of the multijet background
are implemented through a moment morphing technique~\cite{Baak:2014fta}.
Background shape variations are obtained through modifications of the nominal
sampling procedure by shifting the photon \pt and by deforming the
$\Delta\phi(M,\gamma)$ distribution.
These effects are propagated to the $\W$ boson invariant mass shape resulting
in a shift and a skewness variation, respectively.
A third variation is directly obtained through a
multiplicative transformation of the candidate $\W$ boson invariant mass
by a linear function. These variations provide complementary modes of
deformation of the nominal background shape. Each variation is controlled in
the fit by a nuisance parameter.
The pre-fit magnitude of the variation is chosen to be large enough so that the
corresponding parameters are constrained by the data in the fit. Larger
variations were found to produce compatible results.

In the \TRPH category, uncertainties are similar in size for all
three signal processes. The impact of uncertainties in terms of normalization
variation is described in the following.
Trigger efficiency calibration uncertainties are estimated by factorizing the
photon and track components and amount to 0.6\% and 3.6\%, respectively.
The uncertainty for the track component of the trigger is derived by
correcting and smearing the leading track $\ET/\pt$ according to the results
of Ref.~\cite{JETM-2018-08}.
The impact of energy scale and resolution effects~\cite{PERF-2017-03} is found
to be below 1\%. Sources of uncertainty associated with photon identification
and isolation efficiencies~\cite{PERF-2017-02} account for a 2\% normalization
variation, and those associated with track efficiency amount to approximately 1\%.
The estimated uncertainty associated with the correction of the pileup profile
in simulated events is 2.2\%.

In the \TAPH category, the trigger efficiency uncertainty is 10\%,
determined by comparing data and simulated $Z\to\tau (\mu\nu\nu)\tauhad$ events
selected by a muon-photon trigger that uses the same photon selection criteria
used by the di-photon trigger chosen for $\Wrho$.
Energy resolution and scale uncertainties associated with the calorimeter
response amount to 6\%.
A 5.5\% uncertainty is associated with pileup modeling.
The combined impact of reconstruction, identification and isolation efficiency
uncertainties for \tauhad is 13\%, and 2\% for photons.

In both \TRPH and \TAPH categories, signals are subject to
a 3.3\% uncertainty associated with the $pp\to\Wboson$ cross
section~\cite{STDM-2015-03}. The acceptance uncertainty associated with
renormalization and factorization scale variations is estimated conservatively
due to statistical fluctuations as 6.2\% in the \TRPH SR and 6.5\% in the \TAPH SR.
The uncertainty on the integrated luminosity is 0.83\%~\cite{DAPR-2021-01}.

A simultaneous fit is performed including both \TRPH and \TAPH SRs.
The inclusion of both SRs better constrains the \Wrho signal strength
parameter. The \Wpi and \WK contributions in the \TAPH SR are negligible
and not included in the fit.
Uncertainties and background normalization parameters are not correlated across
SRs due to the large difference in phase space, except for those associated
with the \W boson production cross section and the integrated luminosity.

The expected number of \Wpi (\WK) events in the \TRPH SR is
$5.0 \pm 0.4$ ($0.45 \pm 0.04$), assuming the previously quoted branching
fraction $4.0 \times 10^{-9}$ ($3.3 \times 10^{-9}$) and including both
statistical and systematic uncertainties.
The expected number of \Wrho events, with a branching fraction of
$8.7\times10^{-9}$, is
$1.18 \pm 0.10$ in the \TRPH SR and
$0.72 \pm 0.14$ in the \TAPH SR.
The result of a signal-plus-background fit is shown in \cref{fig:fit}, with the
\WM contributions overlaid. The number of observed events is reported in
\cref{tab:fit}.

\begin{figure}[htbp]
\begin{center}
\subfloat[][]{\includegraphics[width=\twofigwidth]{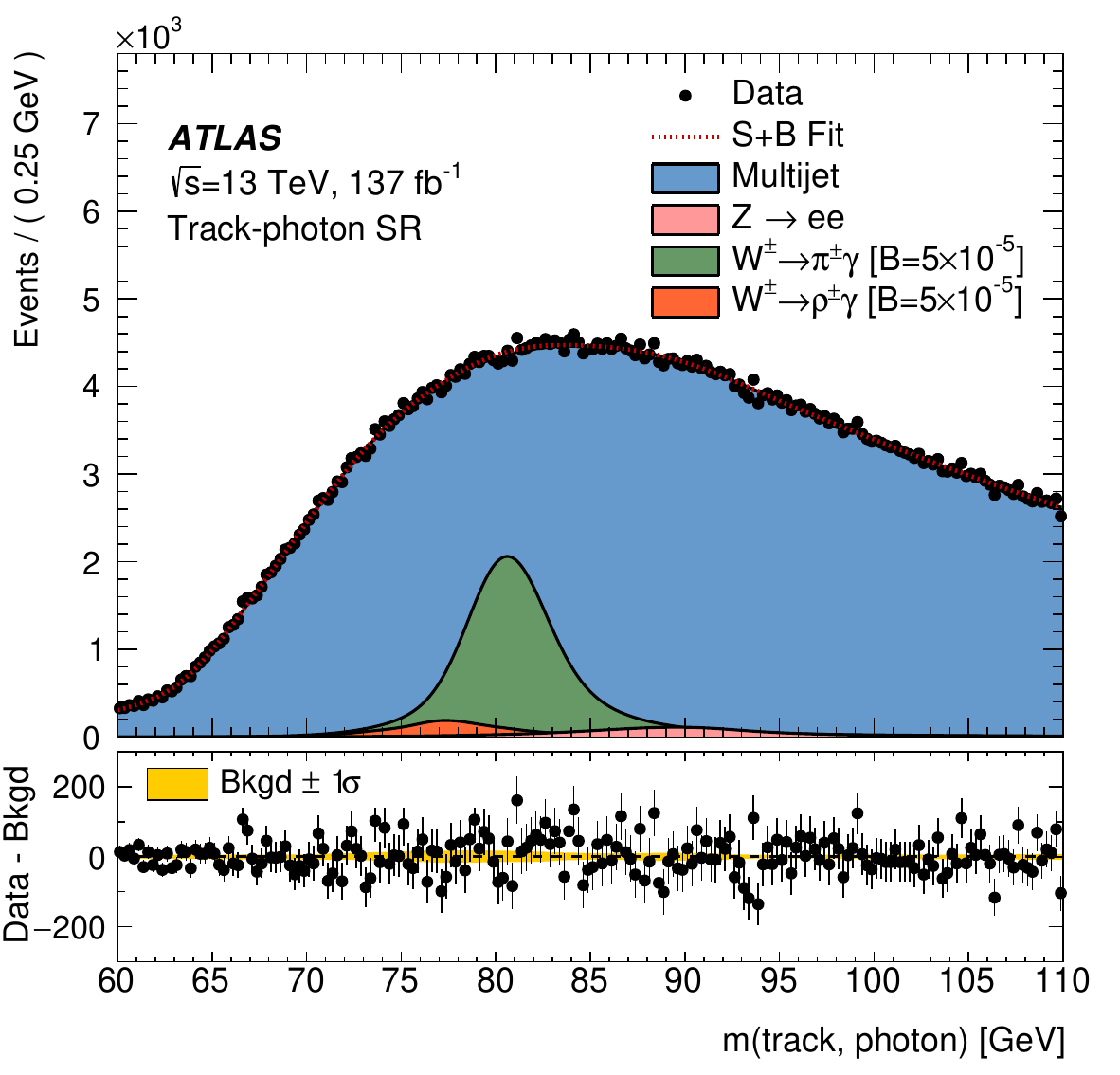}}
\subfloat[][]{\includegraphics[width=\twofigwidth]{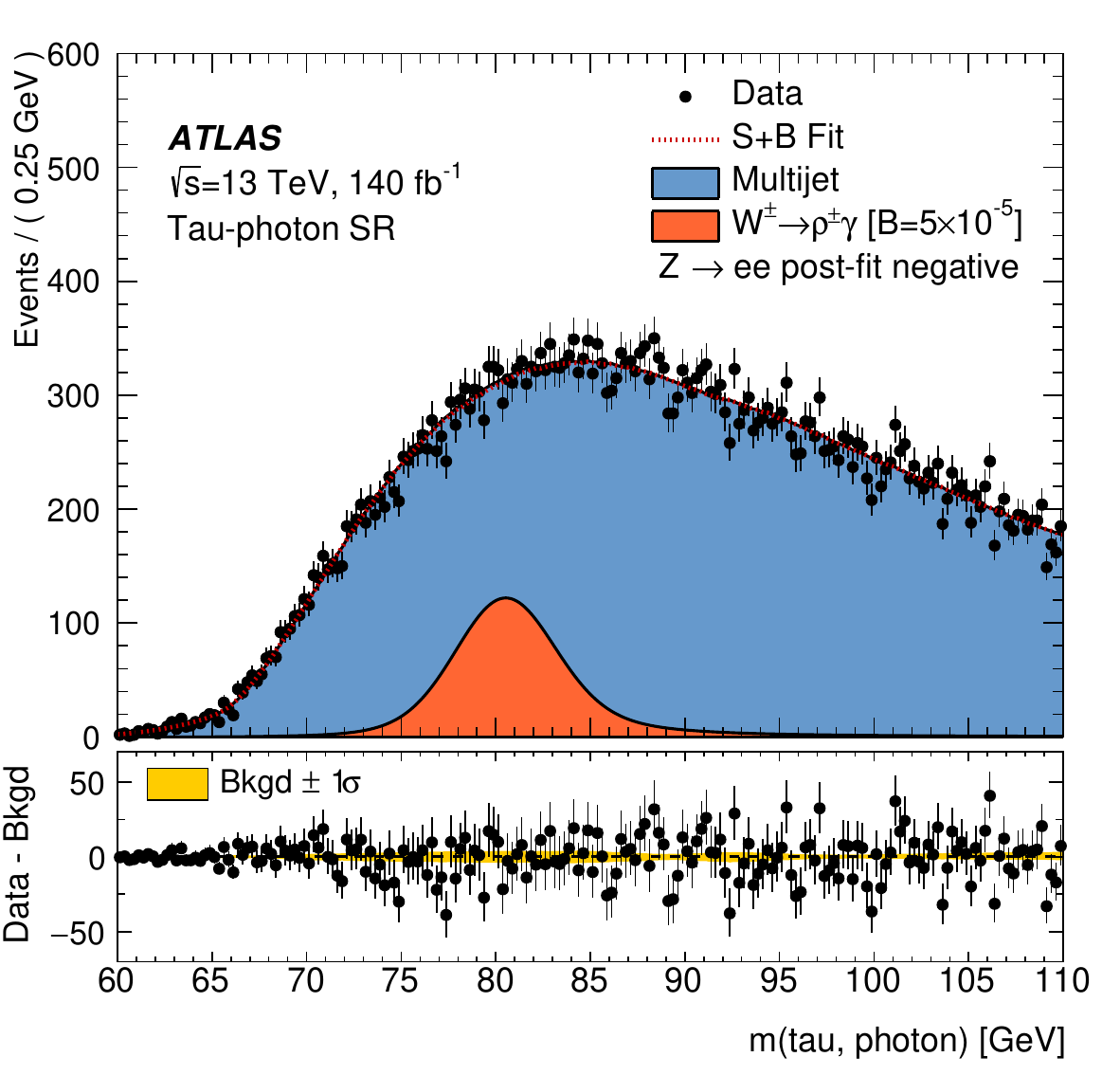}}
\caption{Distributions of the candidate $\Wboson$ boson invariant mass in the
(a) \TRPH SR and (b) \TAPH SR. In the top panel data (black points) are compared
to the signal-plus-background model after a fit.
The $\Wpi$ and $\Wrho$ distributions corresponding to an arbitrarily
large branching fraction of $5\times10^{-5}$ are shown overlaid.
The lower panel displays the difference between the number of data and
background events. The error bars account only for the statistical uncertainty.
The yellow band displays the background systematic uncertainty.}
\label{fig:fit}
\end{center}
\end{figure}

\begin{table}[htpb]
\centering
\caption{Number of expected and observed events in the signal regions extracted
from the signal-plus-background fit. All uncertainties described in the text are
included. \WpiK represents the sum of \Wpi and \WK contributions.}
\label{tab:fit}
\begin{tabular}{lr@{$\ \pm\ $}lr@{$\ \pm\ $}l}
\toprule
& \multicolumn{4}{c}{Number of events} \\
& \multicolumn{2}{c}{Track-photon SR} & \multicolumn{2}{c}{Tau-photon SR} \\
\midrule
Multijet & 632000 & 2200 & 43200  & 600 \\
$\Zee$   & 6100   & 1500 & -200   & 400 \\
$\WpiK$  & 1000   & 800  & \multicolumn{2}{c}{--} \\
$\Wrho$  & -100   & 400  & -90   & 240 \\
Data     & \multicolumn{2}{c}{638962} & \multicolumn{2}{c}{42918} \\
\bottomrule
\end{tabular}
\end{table}

No significant excess with respect to the background prediction is observed in
the data.
Upper limits obtained using the asymptotic approximation of the profile
likelihood test statistic described in Ref.~\cite{Cowan:2010js} and the modified
frequentist confidence level $\text{CL}_{\text{S}}$~\cite{Read:2002hq} are
reported in~\cref{tab:ul}.
When computing the \Wpi upper limit, \Wrho is profiled, and vice-versa.
The \WK upper limit is produced in the same manner, conservatively replacing
\Wpi with \WK.
The systematic uncertainties result in a deterioration of the obtained upper
limit by $+42\%$ for \Wpi and \WK and $+59\%$ for \Wrho. The dominant systematic
uncertainties are the ones associated with the modeling of the shape of the
multijet background: the sole inclusion of signal uncertainties degrades
the upper limit by $+1\%$ for \Wpi and \WK and $+6\%$ for \Wrho.
The combined \TRPH and \TAPH fit improves the observed (expected) upper limit on
\Wrho by $18\%$ (7\%) compared to a \TAPH-only fit. The inclusion of the
\TAPH selection has negligible impact on the \Wpi and \WK upper limits.

\begin{table}[htbp]
\centering
\caption{Expected and observed upper limits on the
$\Wpi$, $\WK$, and $\Wrho$ branching fractions.}
\label{tab:ul}
\begin{tabular}{ccc}
\toprule
& \multicolumn{2}{c}{95\% CL upper limits} \\\cmidrule{2-3}
Branching fraction  & Expected $\times10^{-6}$ & Observed $\times10^{-6}$ \\
\midrule
$\mathcal{B}(\Wpi) $ & $1.2^{+0.5}_{-0.3}$  & 1.9 \\[5pt]
$\mathcal{B}(\WK)  $ & $1.1^{+0.4}_{-0.3}$  & 1.7 \\[5pt]
$\mathcal{B}(\Wrho)$ & $6.0^{+2.3}_{-1.7}$  & 5.2 \\[5pt]
\bottomrule
\end{tabular}
\end{table}

These results improve the previous upper limit on \BWpi~\cite{Aaltonen:2011nn}
by approximately a factor of four and provide first upper limits on \BWK and \BWrho.
This work provides relevant input for the design of future collider experiments,
where exclusive hadronic decays of the \W boson could potentially be observed
for the first time. Future $e^{+}e^{-}$ colliders are expected to deliver a clean
sample of $\mathcal{O}(10^{8})$ $\W^{+}\W^{-}$ events~\cite{Blondel:2021ema},
which would allow access to the \WD and \Wppp decays according to current SM
predictions~\cite{Grossmann:2015lea,Melia:2016knk,Mangano:2014xta,CMS-SMP-18-009}.
Future hadron colliders are projected to produce $\mathcal{O}(10^{12})$ \W
bosons~\cite{Mangano:2022ukr}, which would translate to thousands of \Wpi and
\Wrho decays and hundreds of \WK decays.
In both cases, the observation of these channels pose significant experimental
challenges both in terms of trigger strategy and background discrimination.
Thus, careful detector optimization is required in order to access these
signatures, and exploit them as a new way to measure the \W boson properties and
to probe the QCD factorization formalism.
The novel experimental techniques presented in this Letter are an initial step
towards the observation of these decays in future facilities, which are
currently being planned.


\section*{Acknowledgements}

%
%

%
%

We thank CERN for the very successful operation of the LHC and its injectors, as well as the support staff at
CERN and at our institutions worldwide without whom ATLAS could not be operated efficiently.

The crucial computing support from all WLCG partners is acknowledged gratefully, in particular from CERN, the ATLAS Tier-1 facilities at TRIUMF/SFU (Canada), NDGF (Denmark, Norway, Sweden), CC-IN2P3 (France), KIT/GridKA (Germany), INFN-CNAF (Italy), NL-T1 (Netherlands), PIC (Spain), RAL (UK) and BNL (USA), the Tier-2 facilities worldwide and large non-WLCG resource providers. Major contributors of computing resources are listed in Ref.~\cite{ATL-SOFT-PUB-2023-001}.

We gratefully acknowledge the support of ANPCyT, Argentina; YerPhI, Armenia; ARC, Australia; BMWFW and FWF, Austria; ANAS, Azerbaijan; CNPq and FAPESP, Brazil; NSERC, NRC and CFI, Canada; CERN; ANID, Chile; CAS, MOST and NSFC, China; Minciencias, Colombia; MEYS CR, Czech Republic; DNRF and DNSRC, Denmark; IN2P3-CNRS and CEA-DRF/IRFU, France; SRNSFG, Georgia; BMBF, HGF and MPG, Germany; GSRI, Greece; RGC and Hong Kong SAR, China; ISF and Benoziyo Center, Israel; INFN, Italy; MEXT and JSPS, Japan; CNRST, Morocco; NWO, Netherlands; RCN, Norway; MEiN, Poland; FCT, Portugal; MNE/IFA, Romania; MESTD, Serbia; MSSR, Slovakia; ARIS and MVZI, Slovenia; DSI/NRF, South Africa; MICINN, Spain; SRC and Wallenberg Foundation, Sweden; SERI, SNSF and Cantons of Bern and Geneva, Switzerland; NSTC, Taipei; TENMAK, T\"urkiye; STFC/UKRI, United Kingdom; DOE and NSF, United States of America.

Individual groups and members have received support from BCKDF, CANARIE, CRC and DRAC, Canada; CERN-CZ, PRIMUS 21/SCI/017 and UNCE SCI/013, Czech Republic; COST, ERC, ERDF, Horizon 2020, ICSC-NextGenerationEU and Marie Sk{\l}odowska-Curie Actions, European Union; Investissements d'Avenir Labex, Investissements d'Avenir Idex and ANR, France; DFG and AvH Foundation, Germany; Herakleitos, Thales and Aristeia programmes co-financed by EU-ESF and the Greek NSRF, Greece; BSF-NSF and MINERVA, Israel; Norwegian Financial Mechanism 2014-2021, Norway; NCN and NAWA, Poland; La Caixa Banking Foundation, CERCA Programme Generalitat de Catalunya and PROMETEO and GenT Programmes Generalitat Valenciana, Spain; G\"{o}ran Gustafssons Stiftelse, Sweden; The Royal Society and Leverhulme Trust, United Kingdom.

In addition, individual members wish to acknowledge support from Chile: Agencia Nacional de Investigaci\'on y Desarrollo (FONDECYT 1190886, FONDECYT 1210400, FONDECYT 1230812, FONDECYT 1230987); China: National Natural Science Foundation of China (NSFC - 12175119, NSFC 12275265, NSFC-12075060); Czech Republic: PRIMUS Research Programme (PRIMUS/21/SCI/017); EU: H2020 European Research Council (ERC - 101002463); European Union: European Research Council (ERC - 948254), Horizon 2020 Framework Programme (MUCCA - CHIST-ERA-19-XAI-00), European Union, Future Artificial Intelligence Research (FAIR-NextGenerationEU PE00000013), Italian Center for High Performance Computing, Big Data and Quantum Computing (ICSC, NextGenerationEU), Marie Sklodowska-Curie Actions (EU H2020 MSC IF GRANT NO 101033496); France: Agence Nationale de la Recherche (ANR-20-CE31-0013, ANR-21-CE31-0013, ANR-21-CE31-0022, ANR-22-EDIR-0002), Investissements d'Avenir Idex (ANR-11-LABX-0012), Investissements d'Avenir Labex (ANR-11-LABX-0012); Germany: Baden-Württemberg Stiftung (BW Stiftung-Postdoc Eliteprogramme), Deutsche Forschungsgemeinschaft (DFG - CR 312/5-1); Italy: Istituto Nazionale di Fisica Nucleare (FELLINI G.A. n. 754496, ICSC, NextGenerationEU); Japan: Japan Society for the Promotion of Science (JSPS KAKENHI JP21H05085, JSPS KAKENHI JP22H01227, JSPS KAKENHI JP22H04944); Netherlands: Netherlands Organisation for Scientific Research (NWO Veni 2020 - VI.Veni.202.179); Norway: Research Council of Norway (RCN-314472); Poland: Polish National Agency for Academic Exchange (PPN/PPO/2020/1/00002/U/00001), Polish National Science Centre (NCN 2021/42/E/ST2/00350, NCN OPUS nr 2022/47/B/ST2/03059, NCN UMO-2019/34/E/ST2/00393, UMO-2020/37/B/ST2/01043, UMO-2021/40/C/ST2/00187); Slovenia: Slovenian Research Agency (ARIS grant J1-3010); Spain: BBVA Foundation (LEO22-1-603), Generalitat Valenciana (Artemisa, FEDER, IDIFEDER/2018/048), La Caixa Banking Foundation (LCF/BQ/PI20/11760025), Ministry of Science and Innovation (MCIN \& NextGenEU PCI2022-135018-2, MICIN \& FEDER PID2021-125273NB, RYC2019-028510-I, RYC2020-030254-I, RYC2021-031273-I, RYC2022-038164-I), PROMETEO and GenT Programmes Generalitat Valenciana (CIDEGENT/2019/023, CIDEGENT/2019/027); Sweden: Swedish Research Council (VR 2018-00482, VR 2022-03845, VR 2022-04683, VR grant 2021-03651), Knut and Alice Wallenberg Foundation (KAW 2017.0100, KAW 2018.0157, KAW 2018.0458, KAW 2019.0447); Switzerland: Swiss National Science Foundation (SNSF - PCEFP2\_194658); United Kingdom: Leverhulme Trust (Leverhulme Trust RPG-2020-004); United States of America: U.S. Department of Energy (ECA DE-AC02-76SF00515), Neubauer Family Foundation.

%
%


\clearpage

\section*{Supplemental material}

\subsection*{Full event selection}

The complete set of selection requirements which defines the \TRPH and \TAPH Signal Regions (SR) can be found in~\cref{tab:pi_selection} and~\cref{tab:tau_selection}, respectively. Background enriched regions are defined for use in the background modelling and its validation. The Generation Region (GR) includes the object and event level requirements listed in the tables, as well as the requirements devised in order to suppress the background from $\Zee$ events. Validation regions (VR) are defined by applying one of the SR requirements to the events in the GR, as summarised in the tables.

\begin{table}[htpb]
\centering
\caption{Summary of the \TRPH selection. Photon calorimeter isolation and photon track isolation are imposed by requiring that the sum of transverse energies
of calorimeter clusters not associated to the photon candidate but within a cone of $\dR = 0.4$ is less than $2.45\;\GeV + 0.022 \times \pty$ and the sum of transverse momenta of tracks within a cone of $\dR = 0.2$, excluding possible conversion tracks, is less than 5\% of $ \pty$, respectively. Track isolation is imposed by requiring that the sum of transverse momenta of tracks within $\dR  = 0.2$, excluding the candidate track, is less than 14\% of its $\pt$.}
\begin{tabular}{l}
\toprule
\textbf{Baseline selection}\\
\midrule
\textbf{Photon requirements:}\\
$\pt>30\,\GeV$ or $\pt > 35\,\GeV$, dependent on specific trigger \\
$|\eta|<2.37$, excluding $1.37<|\eta|<1.52$ \\
\textit{Tight} identification~\cite{EGAM-2018-01} \\
\midrule
\textbf{Track requirements:}\\
$\pt>30\,\GeV$, $|\eta|<2.5$ \\
\textit{Tight Primary} identification~\cite{ATL-PHYS-PUB-2015-051}  \\
\midrule
\textbf{Event requirements:}\\
$\eta(M) \times  \eta(\gamma)\geq 0$ if track and $\gamma$ in endcap ($|\eta(M)|>1.5$ and $|\eta(\gamma)|>1.37$)  \\
$\Delta\Phi(M,\gamma) > \pi/2$ \\
\bottomrule
\toprule
\textbf{$\Zee$ suppression requirements:} \\
TRT electron likelihood discriminant $\leq 10$\%~\cite{ATLAS-CONF-2016-024} or \\
TRT electron likelihood discriminant $>10$\% if hadronic leakage $>3$\% \\
\bottomrule
\toprule
\textbf{GR selection:} baseline + $\Zee$ suppression requirements \\
\textbf{VR1 selection:} GR + $\pt(M) >$ 33 \GeV \\
\textbf{VR2a selection:} GR + Photon calorimeter isolation \\
\textbf{VR2b selection:} GR + Photon track isolation\\
\textbf{VR3 selection:} GR  + Track isolation\\
\textbf{SR selection:} all requirements \\
\bottomrule
\end{tabular}
\label{tab:pi_selection}
\end{table}

\begin{table}[htpb]
\centering
\caption{Summary of the \TAPH selection. Photon calorimeter isolation and photon track isolation are imposed by requiring that the sum of transverse energies
of calorimeter clusters not associated to the photon candidate but within a cone of $\dR = 0.4$ is less than $2.45\;\GeV + 0.022 \times \pty$ and the sum of transverse momenta of tracks within a cone of $\dR = 0.2$, excluding possible conversion tracks, is less than 5\% of $ \pty$, respectively.}
\begin{tabular}{l}
\toprule
\textbf{Baseline selection}\\
\midrule
\textbf{Photon requirements:}\\
$\pt>36\,\GeV$ \\
$|\eta|<2.37$, excluding $1.37<|\eta|<1.52$ \\
\textit{Tight} identification \\
Photon calorimeter isolation  \\
Photon track isolation\\
\midrule
\textbf{Tau requirements:}\\
$h^{\pm}\pi^{0}$ decay mode \\
$\pt>26\,\GeV$ \\
$|\eta|<2.5$, $1.37<|\eta|<1.52$ \\
\textit{Medium} identification working point~\cite{ATL-PHYS-PUB-2019-033} \\
\midrule
\textbf{Event requirements:}\\
$\Delta\Phi(\tauhadvis,\gamma) > 2$ \\
\bottomrule
\toprule
\textbf{$\Zee$ suppression requirements:} \\
\textit{Tight} electron-tau discriminant working point~\cite{ATL-PHYS-PUB-2015-045} \\
$E_{T}(\tau)/p_{T}(\text{trk}) > 2.4$ \\
$\dRmax > 0.036$ \\
TRT electron likelihood discriminant $< 0.9$ \\
\bottomrule
\toprule
\textbf{GR selection:} baseline + $\Zee$ suppression requirement \\
\textbf{VR1 selection:} GR +  $\pt(\tauhadvis) > 30\,\GeV$ \\
\textbf{VR2 selection:} GR +  $\dRmax<0.065$ \\
\textbf{VR3 selection:} GR +  $\log(|d_{0}(\tauhadvis)|)<-1.2$ \\
\textbf{SR selection:} all requirements \\
\bottomrule
\end{tabular}
\label{tab:tau_selection}
\end{table}

\clearpage
\subsection*{Multijet background modeling}
Multijet pseudo-events are generated through ancestral sampling using
multi-dimensional PDFs.
Each pseudo-event is fully described by the meson and photon four-vectors and
by the additional variables used in selection criteria applied to GRs that
define SRs.
The sampling procedures used to generate multijet pseudo-events in the \TRPH and
\TAPH selections are detailed below and illustrated in \cref{fig:sampling}.
The candidate meson is described in terms of the object used to reconstruct
it i.e.\ a track and a hadronic $\tau$ lepton in the \TRPH and \TAPH selections,
respectively.
\subsubsection*{Track-photon}
\begin{enumerate}
\item $\pttr$ and $\pty$ are sampled simultaneously from a two-dimensional
template.
\item Track isolation is described in bins of $\pty$ and $\pttr$ using
a three-dimensional template. Given the values sampled in step 1, the
template is projected along the track isolation and a value of
track isolation is randomly sampled.
\item Photon calorimeter isolation is described in bins of $\pty$.
A value for photon calorimeter isolation is sampled from the bin corresponding
to the $\pty$ value sampled in step 1.
\item $\Delta\eta(trk,\gamma)$ and photon track isolation are described in bins
of photon calorimeter isolation using a three-dimensional template.
$\Delta\eta(trk,\gamma)$ and photon track isolation are sampled simultaneously
from the two-dimensional projection corresponding to the photon calorimeter
isolation value obtained in the step 3.
\item $\Delta\phi(trk,\gamma)$ is described in bins of $\Delta\eta(trk,\gamma)$
using a two-dimensional template. Based on the value of  $\Delta\eta(trk,\gamma)$
obtained in step 4, the template is projected along the
$\Delta\phi(trk,\gamma)$ and a value for $\Delta\phi(trk,\gamma)$ is sampled.
\item $\eta(trk)$ and $\phi(trk)$ are sampled independently from the corresponding
one-dimensional templates.
\end{enumerate}
\subsubsection*{Tau-photon}
\begin{enumerate}
\item $\ptta$ is sampled from the corresponding one-dimensional template.
\item $\pty$ is sampled from the distribution obtained projecting the
two-dimensional template of $\pty$, $\ptta$ along $\pty$
using the value of $\ptta$ obtained in step 1.
\item \dRmax, defined as the \dR between the pion track associated to the $\tau$
and the $\tau$ axis, is described in bins of $\ptta$ and $\pty$.
Using the values of $\ptta$ and $\pty$ previously obtained, the three-dimensional
distribution is projected along \dR and a value for \dR is sampled.
\item $\log(|d_{0}(\tau)|)$, where $d_{0}$ is defined as the track transverse impact
parameter of the $\tau$ track, is sampled as a function of $\ptta$ and \dRmax
using the same technique described in step 3.
\item $\eta(\tau)$ is sampled as a function of $\log(|d_{0}(\tau)|)$ and \dRmax using
the same technique described in the previous step.
\item $\Delta\eta(\tau,\gamma)$ is sampled from the projection of the two-dimensional
distribution of $\Delta\eta(\tau,\gamma)$ and $\eta(\tau)$ using $\eta(\tau)$ obtained
in the previous step.
\item $\Delta\phi(\tau,\gamma)$ is sampled from the projection of the
three-dimensional template of $\Delta\phi(\tau,\gamma)$, $\ptta$,
$\pty$ defined by the values previously obtained.
\item $\phi(\tau)$ is sampled independently from the corresponding one-dimensional
template.
\end{enumerate}

\begin{figure}[htbp]
\begin{center}
\subfloat[][]{\includegraphics[width=0.62\textwidth]{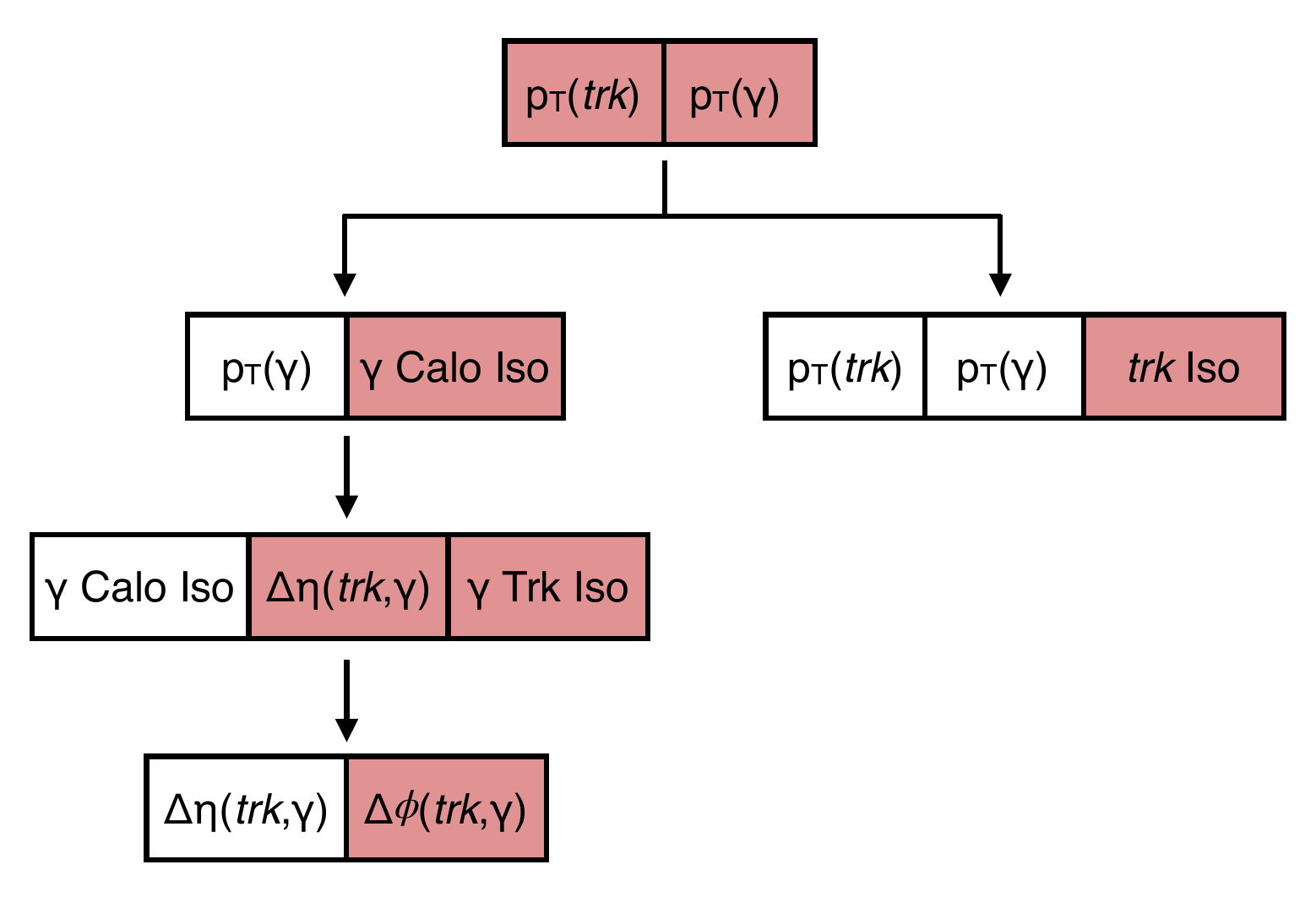}}
\subfloat[][]{\includegraphics[width=0.38\textwidth]{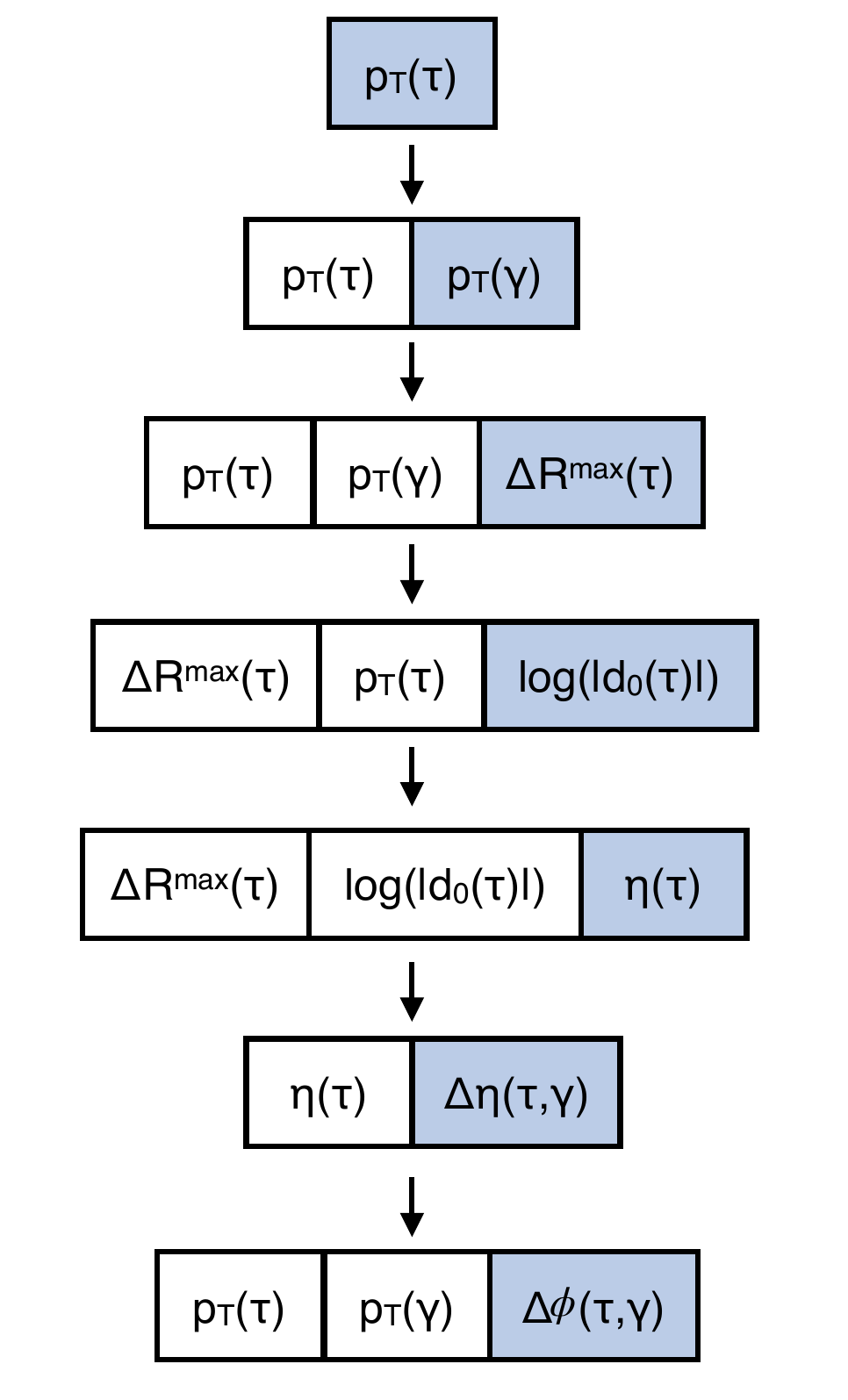}}
\caption{Graphical representation of the sampling sequence used in the generation
of multijet pseudo-events in the (a) \TRPH selection, (b) \TAPH selection.
Variables not shown explicitly are sampled in a factorized, uncorrelated
manner from one-dimensional templates.
Groups of two(three) variables represent two(three)-dimensional templates.
Arrows are used to show the sequential order of steps in the
sampling. Variables are highlighted with color at the step in which they are
defined for each pseudo-candidate.
}
\label{fig:sampling}
\end{center}
\end{figure}


\printbibliography

\clearpage
 
\begin{flushleft}
\hypersetup{urlcolor=black}
{\Large The ATLAS Collaboration}

\bigskip

\AtlasOrcid[0000-0002-6665-4934]{G.~Aad}$^\textrm{\scriptsize 102}$,
\AtlasOrcid[0000-0002-5888-2734]{B.~Abbott}$^\textrm{\scriptsize 120}$,
\AtlasOrcid[0000-0002-1002-1652]{K.~Abeling}$^\textrm{\scriptsize 55}$,
\AtlasOrcid[0000-0001-5763-2760]{N.J.~Abicht}$^\textrm{\scriptsize 49}$,
\AtlasOrcid[0000-0002-8496-9294]{S.H.~Abidi}$^\textrm{\scriptsize 29}$,
\AtlasOrcid[0000-0002-9987-2292]{A.~Aboulhorma}$^\textrm{\scriptsize 35e}$,
\AtlasOrcid[0000-0001-5329-6640]{H.~Abramowicz}$^\textrm{\scriptsize 151}$,
\AtlasOrcid[0000-0002-1599-2896]{H.~Abreu}$^\textrm{\scriptsize 150}$,
\AtlasOrcid[0000-0003-0403-3697]{Y.~Abulaiti}$^\textrm{\scriptsize 117}$,
\AtlasOrcid[0000-0002-8588-9157]{B.S.~Acharya}$^\textrm{\scriptsize 69a,69b,m}$
% ...
\bigskip
\\

% ...
$^{29}$Physics Department, Brookhaven National Laboratory, Upton NY; United States of America.\\
% ...
$^{35}$$^{(a)}$Facult\'e des Sciences Ain Chock, Universit\'e Hassan II de Casablanca;$^{(b)}$Facult\'{e} des Sciences, Universit\'{e} Ibn-Tofail, K\'{e}nitra;$^{(c)}$Facult\'e des Sciences Semlalia, Universit\'e Cadi Ayyad, LPHEA-Marrakech;$^{(d)}$LPMR, Facult\'e des Sciences, Universit\'e Mohamed Premier, Oujda;$^{(e)}$Facult\'e des sciences, Universit\'e Mohammed V, Rabat;$^{(f)}$Institute of Applied Physics, Mohammed VI Polytechnic University, Ben Guerir; Morocco.\\
% ...
$^{49}$Fakult\"{a}t Physik , Technische Universit{\"a}t Dortmund, Dortmund; Germany.\\
% ...
$^{55}$II. Physikalisches Institut, Georg-August-Universit\"{a}t G\"ottingen, G\"ottingen; Germany.\\
% ...
$^{69}$$^{(a)}$INFN Gruppo Collegato di Udine, Sezione di Trieste, Udine;$^{(b)}$ICTP, Trieste;$^{(c)}$Dipartimento Politecnico di Ingegneria e Architettura, Universit\`a di Udine, Udine; Italy.\\
% ...
$^{102}$CPPM, Aix-Marseille Universit\'e, CNRS/IN2P3, Marseille; France.\\
% ...
$^{117}$Department of Physics, New York University, New York NY; United States of America.\\
% ...
$^{120}$Homer L. Dodge Department of Physics and Astronomy, University of Oklahoma, Norman OK; United States of America.\\
% ...
$^{150}$Department of Physics, Technion, Israel Institute of Technology, Haifa; Israel.\\
$^{151}$Raymond and Beverly Sackler School of Physics and Astronomy, Tel Aviv University, Tel Aviv; Israel.\\
% ...
$^{a}$ Also Affiliated with an institute covered by a cooperation agreement with CERN.\\
$^{b}$ Also at An-Najah National University, Nablus; Palestine.\\
$^{c}$ Also at Borough of Manhattan Community College, City University of New York, New York NY; United States of America.\\
$^{d}$ Also at Center for High Energy Physics, Peking University; China.\\
$^{e}$ Also at Center for Interdisciplinary Research and Innovation (CIRI-AUTH), Thessaloniki; Greece.\\
$^{f}$ Also at Centro Studi e Ricerche Enrico Fermi; Italy.\\
% ...
$^{m}$ Also at Department of Physics, King's College London, London; United Kingdom.\\
% ...

\end{flushleft}


%

%
%

%

%
%
%
%

\end{document}